\documentclass[11pt]{article}

\renewcommand{\arraystretch}{1.3}
\makeatletter
\newdimen\normalarrayskip              
\newdimen\minarrayskip                 
\normalarrayskip\baselineskip \minarrayskip\jot
\newif\ifold             \oldtrue            \def\new{\oldfalse}
\def\arraymode{\ifold\relax\else\displaystyle\fi} 
\def\eqnumphantom{\phantom{(\theequation)}}     
\def\@arrayskip{\ifold\baselineskip\z@\lineskip\z@
     \else
     \baselineskip\minarrayskip\lineskip2\minarrayskip\fi}
\def\@arrayclassz{\ifcase \@lastchclass \@acolampacol \or
\@ampacol \or \or \or \@addamp \or
   \@acolampacol \or \@firstampfalse \@acol \fi
\edef\@preamble{\@preamble
  \ifcase \@chnum
     \hfil$\relax\arraymode\@sharp$\hfil
     \or $\relax\arraymode\@sharp$\hfil
     \or \hfil$\relax\arraymode\@sharp$\fi}}
\def\@array[#1]#2{\setbox\@arstrutbox=\hbox{\vrule
     height\arraystretch \ht\strutbox
     depth\arraystretch \dp\strutbox
     width\z@}\@mkpream{#2}\edef\@preamble{\halign
\noexpand\@halignto
\bgroup \tabskip\z@ \@arstrut \@preamble \tabskip\z@ \cr}%
\let\@startpbox\@@startpbox \let\@endpbox\@@endpbox
  \if #1t\vtop \else \if#1b\vbox \else \vcenter \fi\fi
  \bgroup \let\par\relax
  \let\@sharp##\let\protect\relax
  \@arrayskip\@preamble}
\def\eqnarray{\stepcounter{equation}%
              \let\@currentlabel=\theequation
              \global\@eqnswtrue
              \global\@eqcnt\z@
              \tabskip\@centering
              \let\\=\@eqncr
 \halign to \displaywidth\bgroup
    \eqnumphantom\@eqnsel\hskip\@centering
    $\displaystyle \tabskip\z@ {##}$%
    \global\@eqcnt\@ne \hskip 2\arraycolsep
         $\displaystyle\arraymode{##}$\hfil
    \global\@eqcnt\tw@ \hskip 2\arraycolsep
         $\displaystyle\tabskip\z@{##}$\hfil
         \tabskip\@centering
    &{##}\tabskip\z@\cr}
\begingroup\ifx\undefined\newsymbol \else\def\input#1 {\endgroup}\fi

\catcode`\@=11
\def\marginnote#1{}

\newcount\hour
\newcount\minute
\newtoks\amorpm
\hour=\time\divide\hour by60 \minute=\time{\multiply\hour by60
\global\advance\minute by-\hour}
\edef\standardtime{{\ifnum\hour<12 \global\amorpm={am}%
        \else\global\amorpm={pm}\advance\hour by-12 \fi
        \ifnum\hour=0 \hour=12 \fi
        \number\hour:\ifnum\minute<10 0\fi\number\minute\the\amorpm}}
\edef\militarytime{\number\hour:\ifnum\minute<10 0\fi\number\minute}

\def\draftlabel#1{{\@bsphack\if@filesw {\let\thepage\relax
      \xdef\@gtempa{\write\@auxout{\string
          \newlabel{#1}{{\@currentlabel}{\thepage}}}}}\@gtempa \if@nobreak
    \ifvmode\nobreak\fi\fi\fi\@esphack} \gdef\@eqnlabel{#1}}
    \def\@eqnlabel{}
\def\@vacuum{}
\def\draftmarginnote#1{\marginpar{\raggedright\scriptsize\tt#1}}

\def\draft{
  \oddsidemargin -.5truein
  \def\@oddfoot{\footnotesize \sl preliminary draft \hfil
    \rm\thepage\hfil\sl\today\quad\militarytime}
  \let\@evenfoot\@oddfoot \overfullrule 3pt
    \let\label=\draftlabel
    \let\marginnote=\draftmarginnote
  \def\@eqnnum{(\theequation)\rlap{\kern\marginparsep\tt\@eqnlabel}%
    \global\let\@eqnlabel\@vacuum}

  }

\def\be{\begin{eqnarray}}
\def\ee{\end{eqnarray}}
\def\nn{\nonumber}

\def\beq{\begin{equation}}
\def\eeq{\end{equation}}
\def\ba{\beq\new\begin{array}{c}}
\def\ea{\end{array}\eeq}
\def\be{\ba}
\def\ee{\ea}

\newfont{\alef}{msbm10 at 12pt}
\newfont {\goth}{eufm10 at 11pt}
\def\mathbb#1{\hbox{{\alef #1}}}

\let\@@savethanks\thanks
\def\thanks#1{\gdef\thefootnote{\alph{footnote}}\@@savethanks{#1}}

\def\theequation{\arabic{section}.\arabic{equation}}

\title{{\bf On the Problem of Radiation Friction Beyond 4 and 6
Dimensions } \vspace{.5cm}}
\author{{\bf A. Mironov}\footnote{E-mail: \ mironov@itep.ru; mironov@lpi.ru}
\date{ } \\
{\small {\it Lebedev Physics Institute}
and {\it ITEP, Moscow, Russia}}\\ \\
{\bf A. Morozov}\thanks{E-mail: \ morozov@itep.ru}
\date{ } \\ {\small {\it ITEP, Moscow, Russia}}}

\begin{document}

\maketitle

\vspace{-8.5cm}

\begin{center}
\hfill FIAN/TD-21/07\\
\hfill ITEP/TH-42/07\\
\end{center}

\vspace{7.0cm}

\begin{abstract}
\noindent We count the number of independent structures which can
arise in expressions for radiation friction force in different even
space-time dimensions and demonstrate that their number is too big
at $d\geq 8$ to allow determination of this force from the
transversality condition alone, as was done by B.Kosyakov in $6d$.
This implies that in general one can not bypass a tedious
calculation involving explicit regularization and evaluation of
emerging counterterms. However, simple Kosyakov's method works
nicely in any dimension for the special case of circular motion with
constant angular velocity.
\end{abstract}

\section{Introduction}

If radiation carries away the energy-momentum from a point like
source
at the rate $W_\mu$, 
then the radiation friction force $F_\mu$ should appear at the
r.h.s. of source's equation of motion, \be m\dot u_\mu = f_\mu \ \
\longrightarrow \ \ m\dot u_\mu = f_\mu + F_\mu \label{eqm} \ee
where $u_\mu = \gamma(1,\vec v)$, $\gamma = (1-\vec v^2)^{-1/2}$ is
relativistic velocity of the source, dot denotes derivative w.r.t.
the particle self-time $\tau$ and $f_\mu$ is the relativistic force,
which causes source's acceleration.

Energy-momentum conservation implies that the work of the radiation
friction force should compensate the energy-momentum outflow \be
F_\mu = W_\mu 
+ \dot\xi_\mu, \ee where the last term at the r.h.s. takes into
account the change of the energy-momentum of electromagnetic field
in the "near domain" around the source, which is not carried away to
infinity, and is a total $\tau$-derivative of an expression
$\xi_\mu$, made from velocity $u$ and its $\tau$-derivatives.

Since
\be
u^2=1
\label{u2}
\ee
the l.h.s. of eq.(\ref{eqm}) is orthogonal to $u$,
\be
u\dot u = u^\mu \dot u_\mu = 0,
\label{uudot}
\ee
and so are relativistic forces at the r.h.s.,
$u^\mu f_\mu = 0$ and
\be
u^\mu F_\mu = 0
\label{ortho}
\ee

Evaluation of the radiation friction is a tricky task, far more
sophisticated than that of $W_\mu$, because it requires separation
between the fields in the "near" and "far" (wave) domains and also
involves discussion of interaction between the charge and its own
field, related to celebrated problems like electromagnetic mass and
Poincare tension. Still the force itself is a well defined -- and
even experimentally measurable -- quantity, and one can be
interested in knowing the answer for it irrespective of the details
of the deep theory. Thus it is natural to search for
short-cut ways to calculate radiation friction.

Note that our discussion of the radiation friction is formally
applicable equally well for radiation of any spin $s$. However, for
$s>1$, the energy-momentum tensor for the point-like source is not
conserved. This usually means that one cannot neglect contributions
to radiation from tensions of the forces that cause acceleration of
the source. Ultimately, it leads to additional contributions into
the radiation friction.

\section{Kosyakov's trick}

The simplest and the most elegant option \cite{Kos} is to take the
well-known expression for $W_\mu$ \cite{grarad} and to {\it
construct} an expression for $\xi_\mu(u,\dot u, \ldots)$, which
satisfies orthogonality condition (\ref{ortho}), i.e. adjust
$\xi_\mu$ to satisfy \be u^\mu \dot\xi_\mu + u^\mu W_\mu=0
\label{ortho2} \ee Unfortunately, as explained below in this
section, this trick, while effective in $4$ and even in $6$
dimensions, appears non-applicable in general, for $d\geq 8$.

Let $u_{kl} = \partial_\tau^k u^\mu \partial_\tau^l u_\mu$
denote scalar bilinears in $\tau$-derivatives of $u$.
Because of (\ref{u2}) they are not all independent:
$u_{0m}$ can be expressed through $u_{k,m-k}$ with
$1\leq k \leq l-1$:
\be
u_{01} = 0, \ \ {\rm see}\ (\ref{uudot}), \nn \\
u_{02} = - u_{11}, \nn \\
u_{03} = -3u_{12}, \nn \\
u_{04} = - 3u_{22}-4u_{13}, \nn \\
u_{05} = -10u_{23}-5u_{14}, \nn \\
u_{06} = - 10 u_{33}-15 u_{24} - 6u_{15}, \nn \\
\ldots \nn \\
u_{0m} = -\sum_{k=1}^{\left[\frac{m-1}{2}\right]}
C_m^k u_{k,m-k} - \frac{1}{2} C_m^{m/2}\ u_{m/2,m/2}
\label{u0m}
\ee
The last term in the last formula is present only for even $m$,
square brackets in the upper summation limit in the previous term
denote integer part of $\frac{m-1}{2}$. Note in passing that the
total number of items in the relation for $u_{0m}$ is $2^{m-1}$;
this is because they all arise as multiple derivatives of
(\ref{u2}).

Expression for the radiated energy-momentum in $d$ space-time
dimensions ($d$ even) looks as follows \cite{MM2}: \be W_\mu^{(d)} =
\sum_{m=0}^{d-4} w_m^{(d)} \partial_\tau^m u_\mu \label{Wexpan} \ee
where $w_m^{(d)}$ is a polynomial in $u_{kl}$, a linear combination
of $N_{d-2-m}$ monomials $u_{k_1l_1}\ldots u_{k_rl_r}$ with any $r$
(actually, $r \leq \frac{d-2-m}{2}$) and parameters $k_i,l_i,\ i
=1\ldots r$ constrained by the conditions \be 1\leq k_i \leq l_i
\label{consneq} \ee and \be \sum_{i=1}^r (k_i+l_i) = d-2-m.
\label{sum} \ee
On dimensional grounds also a term with
$\partial_\tau^{d-2} u_\mu$ is allowed, but actually it
does not contribute to $W_\mu^{(d)}$: the coefficient
\be
w_{d-2}^{(d)} = 0
\ee
There is no such a restriction in the case of $\xi_\mu^{(d)}$,
instead it has one dimension less, and \be \xi_\mu^{(d)} =
\sum_{m=0}^{d-3}
\kappa_m^{(d)} \partial_\tau^m u_\mu
\label{xiexpan}
\ee
with polynomials $\kappa$ 
are similar to $w$, only this time $\sum_{i=1}^r (k_i+l_i) = d-3-m$.
It follows that the scalars $u^\mu W_\mu$
and $u^\mu \dot\xi_\mu$ are polynomials
in $u_{kl}$ with the total number $\sum (k+l) = d-2$.

In order to check applicability of Kosyakov's trick we need to
compare the two numbers: \be \tilde N_{d-2} = \sum_{m=0}^{d-3}
N_{d-3-m} \label{tildeNvsN} \ee and $N_{d-2}$: the first one counts
the total number of monomials, that {\it can\ } contribute to
$\xi_\mu^{(d)}$, and the second one counts the number of monomials
that can appear in $u^\mu W_\mu$ and which $\xi_\mu$ is supposed to
compensate for according to (\ref{ortho2}). The trick works if
$\tilde N_{d-2} = N_{d-2}$: then the condition (\ref{ortho2}) allows
one to {\it unambiguously\ } extract all the coefficients in
(\ref{xiexpan}) from (\ref{ortho2}) -- and this is indeed the case
for $d=4$ and $d=6$. Unfortunately, for higher dimensions the
matching breaks down: $\tilde N_{d-2} > N_{d-2}$ for $d\geq 8$, and
$\xi_\mu$ is only constrained by (\ref{ortho2}), some $\tilde
N_{d-2} - N_{d-2}$ coefficients in $\xi_\mu^{(d)}$ remain undefined,
and other methods should be used in order to fix them unambiguously.

The numbers $N_k$ are close relatives of the numbers $n_k$
which count natural partitions of $k$,
\be
1 + \sum_{k=1}^\infty n_kq^k =
\prod_{m=1}^\infty \frac{1}{1-q^m},
\label{eta}
\ee
however, $n_k$ would take only (\ref{sum}) into account,
while in the case of $N_k$ an additional constraint
(\ref{consneq}) is imposed
and members of every partition should be grouped in pairs:
$$
\begin{array}{lcc}
1  & n_1 = 1 & N_1 = 0 \\
2 = \underline{1+1} & n_2 = 2 & N_2 = 1 \\
3 = 1 + 1 + 1 = \underline{1 + 2}  & n_3=3 & N_3 = 1 \\
4 = \underline{1+1+1+1} = 1+1+2 = \underline{1+3} =
\underline{2+2} & n_4 = 5 & N_4 = 3 \\
5 = 1+1+1+1+1 = \underline{1+1+1+2} = 1+1+3 =
\underline{1+4} = 1+2+2 = \underline{2+3} \ \ \
& n_5 = 7 & N_5=3\\
6 = \underline{1+1+1+1+1+1} = 1+1+1+1+2 = \underline{1+1+1+3}
= 1+1+4 = & & \\
\ \ \ \ \
= \underline{1+5} = \underbrace{1+1+2+2}_2 = 1+2+3 = 2+2+2 =
\underline{2+4} = \underline{3+3} & n_6 = 11 & N_6 = 7 \\
&\ldots &
\end{array}
$$
Underlined and under-braced are partitions, contributing to
$N_k$, under-braced are partitions which contribute
{\it several\ } times to $N_k$.

Generating function of numbers $N_k$ is given by
\be
{\cal N}(q) = \sum_{k=0}^\infty N_k q^k =
\sum_{r=0}^\infty {\cal N}_r(q)
\ee
where ${\cal N}_r(q)$ counts the numbers of polynomials
of degree $r$ in $u_{kl}$.
We have:
$$
{\cal N}_0(q) = 1,
$$
$$
{\cal N}_1(q) = q(q+q^2+q^3+\ldots) + q^2(q^2+q^3+q^4+\ldots) +
q^3(q^3+q^4+\ldots) + \ldots = \frac{q^2}{(1-q)(1-q^2)} =
$$
\be
=  \sum_{k=1}^\infty n_{1k}q^k = q^2+q^3 +
2q^4+2q^5+3q^6+3q^7+4q^8+4q^9 + \ldots =
\sum_{k=1}^\infty k\left(q^{2k}+q^{2k+1}\right),
\label{calN1}
\ee
$$
{\cal N}_2(q) =
\frac{1}{2}\Big({\cal N}_1^2(q) + {\cal N}_1(q^2)\Big)
= \frac{q^4(1+q^3)}{(1-q)(1-q^2)^2(1-q^4)} =
$$
$$
= q^4 + q^5+3q^6 + 4q^7 + 8q^8 + 10q^9+16 q^{10} + \ldots,
$$
$$
{\cal N}_3(q) =
\frac{1}{6}\Big({\cal N}_1^3(q) +
3{\cal N}_1(q^2){\cal N}_1(q) + 2{\cal N}_1(q^3)\Big) 
$$
$$
= q^6 + q^7 + 3q^8 + \ldots,
$$
$$
{\cal N}_4(q) =
\frac{1}{24}\Big({\cal N}_1^4(q) +
6{\cal N}_1(q^2){\cal N}_1^2(q) + 3{\cal N}_1^2(q^2)
+ 8{\cal N}_1(q^3){\cal N}_1(q) + 6{\cal N}_1(q^4)\Big)
$$
and so on.
Collecting all terms, we obtain:
$$
{\cal N}(q) =
\sum_{r=0}^\infty {\cal N}_r(q) =
\exp\left({\cal N}_1(q) + \frac{1}{2}{\cal N}_1(q^2) +
\frac{1}{3}{\cal N}_1(q^3) + \ldots\right) =
\exp\left( \sum_{k=1}^\infty \frac{1}{k}{\cal N}_1(q^k)\right)=
$$
\be \stackrel{(\ref{calN1})}{=}\ \prod_{k=1}^\infty
\frac{1}{(1-q^k)^{n_{1k}}} =\prod_{k=1}^\infty
\frac{1}{(1-q^{2k})^k(1-q^{2k+1})^k} \ee The product
$\prod_{k=1}^\infty (1-q^{2k})^{-k}$ is familiar from the theory of
$3d$ partitions (McMahon), it plays the same role as
$\prod_{k=1}^\infty (1-q^k)^{-1}$ from (\ref{eta}) for ordinary
partitions. According to (\ref{tildeNvsN}) the generating function
for the numbers $\tilde N_q$ is equal to \be \tilde{\cal N}(q) =
\frac{q{\cal N}(q)}{1-q} \ee

The first few numbers $N_k$ are given in the following table,
where also the relevant monomials are explicitly listed:

\bigskip

$$
\begin{array}{|c|c|c|}
\hline
&&\\
k & N_k & {\rm monomials} \\
&&\\
\hline
&&\\
0 & 1 & 1 = u^2 = u_{00} \\
&&\\
\hline
&&\\
1 & 0 & 0 = (u\dot u) = u_{01} \\
&&\\
\hline
&&\\
2 & 1 & u_{11} = \dot u^2 \\
&&\\
\hline
&&\\
3 & 1 & u_{12} \\
&&\\
\hline
&&\\
4 & 3 & u_{11}^2, \ u_{13},\ u_{22} \\
&&\\
\hline
&&\\
5 & 3 & u_{11}u_{12},\ u_{14}, u_{23} \\
&&\\
\hline
&&\\
6 & 7 & u_{11}^3,\ u_{12}^2, u_{11}u_{13}, u_{11}u_{22},\
u_{15}, u_{24}, u_{33} \\
&&\\
\hline
&&\\
7 & 8 & u_{11}^2u_{12},\ u_{11}u_{14}, u_{11}u_{23},
u_{12} u_{13}, u_{12}u_{22},\ u_{16}, u_{25}, u_{34} \\
&&\\
\hline
&&\\
8 & 16 & u_{11}^4,\ u_{11}^2 u_{13}, u_{11}^2u_{22},
u_{12}^2u_{11},\
u_{11}u_{15}, u_{11}u_{24}, u_{11}u_{33},
u_{12}u_{14}, u_{12}u_{23}, u_{13}^2,
u_{13}u_{22}, u_{22}^2,\ u_{17}, u_{26}, u_{35}, u_{44}\\
&&\\
\hline
\end{array}
$$

\bigskip

It is now easy to find $\tilde N_{d-2}$ and compare them
with $N_{d-2}$:

\bigskip

$$
\begin{array}{|c||c|c|c|c|c|c|c|c|}
\hline
&&&&&&&&\\
d && 4 && 6 && 8 && 10 \\
&&&&&&&&\\
\hline
&&&&&&&&\\
k = d-2 & 1 & 2 & 3 & 4 & 5 & 6 & 7 & 8 \\
&&&&&&&&\\
\hline
&&&&&&&&\\
N_k & 0 & 1 & 1 & 3 & 3 & 7 & 8 & 16 \\
&&&&&&&&\\
\hline
&&&&&&&&\\
\tilde N_k & 1 & 1 & 2 & 3 & 6 & 9 & 16 & 24 \\
&&&&&&&&\\
\hline
\end{array}
$$

\bigskip

For $d=4$ we have the exact matching, $\tilde N_2 = N_2 = 1$, in
more details \cite{grarad,MM2} (here $s$ is the spin of radiation)
\be
W^{(4)}_\mu = -{4-12s\over 3}\pi u_\mu u_{11}, \nn \\
u^\mu W^{(4)}_\mu = -{4-12s\over 3}\pi u_{11} \ \ - \ {\rm only\
one\ term,\ because}\ N_2 = 1,
\nn \\
\xi^{(4)}_\mu = -{4-12s\over 3}\pi \dot u_\mu \ \ - \ {\rm only\
one\ term,\ because}\ \tilde N_2 = 1,
\nn \\
F^{(4)}_\mu = W^{(4)}_\mu + \dot \xi^{(4)}_\mu = -{4-12s\over 3}\pi
\Big(\ddot u_\mu + u_\mu u_{11}\Big) \ee In the non-relativistic
limit, the first term dominates and \be \vec F^{(4)} \approx
-{4-12s\over 3}\pi \ddot {\vec v}. \ee

Similarly, for $d=6$ we also have the exact matching, $\tilde N_4 =
N_4 = 3$, in more details \cite{Kos,MM2} \be W^{(6)}_\mu =
u_\mu\Big(\alpha u_{22} + \beta u_{11}^2\Big) + \gamma\dot u_\mu
u_{12} + \delta\ddot u_\mu u_{11}, \nn \\
u^\mu W^{(6)}_\mu = \alpha u_{22}+(\beta-\delta)u_{11}^2 \ \ - \
{\rm three\ terms\ contribute,\ because}\ N_3 = 1,
\nn \\
\xi^{(6)}_\mu = -4\alpha u_\mu u_{12} + (\beta-\delta) \dot u_\mu
u_{11} -\alpha \dot{\ddot u}_\mu \ \ - \ {\rm three\ terms\
contribute,\ because} \ \tilde N_4 = 1,
\nn \\
F^{(6)}_\mu = W^{(6)}_\mu + \dot \xi^{(6)}_\mu = -\alpha\ddot{\ddot
u}_\mu + \beta \ddot u_\mu u_{11} + (\gamma-4\alpha+2\beta+2\delta)
\dot u_\mu u_{12} + u_\mu\Big(\beta u_{11}^2-3\alpha u_{22} -4\alpha
u_{13}\Big) \ee The coefficients here are equal to \be
\alpha={8\pi^2\over15}(1-5s),\ \ \ \
\beta= \pi^2\left[{19\over 3}-(2s-3)^2\right]\\
\gamma= {16\pi^2\over 35}(2-7s),\ \ \ \  \delta= {16\pi^2\over
105}(7s-4) \ee

In the non-relativistic limit, the first term dominates (this is the
case in all dimensions!), and \be \vec F^{(6)} \approx -\alpha
\ddot{\ddot{\vec v}}={8\pi^2\over 15}(5s-1)\ddot{\ddot{\vec v}} \ee

In fact, in the expressions above the sign of $W_{\mu}$ was chosen
so that it describes correctly all $s$ but $s=0$ (scalar). In the
latter case, one should reverse the sign. This is because in all but
the scalar cases only the spatial components of all non-zero spin
fields have any physical meaning (e.g. survive in physical gauges),
thus giving rise to the overall minus sign of the kinetic part of
the energy-momentum tensor as compared with the scalar case. This
means that one should also reverse the sign of $\xi_\mu$ and $\vec
F$. Let us stress again that the results for only $s=0,1$ (scalar
and electromagnetic radiations) have practical applicability (see
the Introduction).

Unfortunately, already for $d=8$ the matching fails, $\tilde N_6 =
9\
>\ N_6 = 7$. Mismatch does not allow one to define the coefficients
in front of $2 = \tilde N_6 - N_6$ structures, which can potentially
contribute to $\xi^{(8)}_\mu$, but remain unconstrained by
(\ref{ortho2}). These two structures are: \be \zeta_\mu^{(8,1)} =
3u_\mu u_{11}u_{12} + \dot u_\mu (u_{13}+u_{22}) + 2\ddot u_\mu
u_{12} \label{zeta81} \ee and \be \zeta_\mu^{(8,2)} = 3u_\mu
u_{11}u_{12} - \dot u_\mu u_{13} + \dot{\ddot u}_\mu u_{11}
\label{zeta82} \ee -- it is easy to check that the $\tau$-derivative
of any of the two is orthogonal to $u^\mu$: \be u^\mu\dot\zeta_\mu
\equiv  0. \ee

\section{Circular motion}

The above calculus becomes somewhat different in the special when
the source moves along a circular orbit with constant value of
velocity, namely when $\vec v^2$ and thus $u_0 = \gamma$ do not
change with time. The spatial vector $\vec u = \gamma \vec v$
changes direction, but the acceleration is orthogonal to the
velocity, $\vec v \dot{\vec v} = 0$, and, as a corollary, all \be
u_{kl} = 0\ \ \ {\rm if}\ \ \ k+l \ \ \ {\rm is\ odd}
\label{ukloddcirc} \ee Taking time-derivative of these relations, we
obtain
\be
u_{k,l+1} + u_{k+1,l} = 0\ \ \
{\rm if}\ \ \ k+l+1 \ \ \ {\rm is\ even},
\label{uklevcirc}
\ee
for example, $u_{13}+u_{22} = \dot u_{12} = 0$.
In addition,
\be
\partial_\tau^k u_\mu \partial_\tau^l u_\nu =
\partial_\tau^k u_\mu \partial_\tau^l u_\nu
\ \ \ {\rm for} \ \ k<l \ \ {\rm and\ even\ difference} \ \ l-k,
\label{ukul} \ee for example, $\dot u_\mu \dot{\ddot u}_\nu = \dot
u_\nu \dot{\ddot u}_\mu$. An immediate corollary of relations
(\ref{ukloddcirc})-(\ref{ukul}) is that both $\zeta_\mu^{(8,1)} =
\zeta_\mu^{(8,2)} = 0$. This means that {\it for circular motion}
there is no uncertainty and Kosyakov's trick is sufficient to
determine the radiation friction unambiguously, at least, for $d=8$.

In fact, this is true also for $d=10$ and, moreover, for all even
dimensions $d$. In general, we have for circular motion with
constant velocity:

\bigskip

$$
\begin{array}{|c|c|c|c|c|}
\hline
&&&&\\
d& k = d-2 &  {\rm monomials}
& N^{circ}_k & \tilde N^{circ}_k \\
&&&&\\
\hline
&&&&\\
2 & 0 &  1 = u^2 = u_{00} & 1 &   \\
&&&&\\
\hline
&&&&\\
 & 1  & 0 = (u\dot u) = u_{01} & 0 & \\
&&&&\\
\hline
&&&&\\
4 & 2  & u_{11} = \dot u^2 & 1 & 1 \\
&&&&\\
\hline
&&&&\\
 & 3  & u_{12} & 0 & \\
&&&&\\
\hline
&&&&\\
6 & 4  & u_{11}^2, \ u_{13}=-u_{22}  & 2 & 2\\
&&&&\\
\hline
&&&&\\
& 5  & u_{11}u_{12} = u_{14}= u_{23} = 0 & 0 &  \\
&&&&\\
\hline
&&&&\\
8 & 6  & u_{11}^3,\  u_{11}u_{13}=- u_{11}u_{22},\
u_{15}=-u_{24}=u_{33};\ \ u_{12}^2=0, & 3 & 4-1=3 \\
&&&&\\
\hline
&&&&\\
& 7  & u_{11}^2u_{12} = u_{11}u_{14} = u_{11}u_{23} =
u_{12} u_{13}= u_{12}u_{22} = u_{16} = u_{25} = u_{34}=0
& 0 &  \\
&&&&\\
\hline
&&&&\\
10 & 8  & u_{11}^4,\ u_{11}^2 u_{13} = -u_{11}^2u_{22},\
u_{11}u_{15}=-u_{11}u_{24} =u_{11}u_{33}\ [=]\
u_{13}^2=-u_{13}u_{22}=u_{22}^2,
& 5-1 & 7-3 \\
&&
u_{17}=-u_{26}=u_{35}=-u_{44};\ \
u_{12}^2u_{11}=0,\
u_{12}u_{14}=u_{12}u_{23}=0,\
 &=4&=4\\
&&&&\\
\hline
\end{array}
$$

\bigskip

Numbers $N_k^{circ}$ differ from $N_k$ in the previous tables,
because of relations (\ref{ukloddcirc}) and (\ref{uklevcirc}). Only
{\it bilinear} combinations of the last relations (\ref{ukul})
affect $N_k^{circ}$, and this happens for the first time for $k=8$,
i.e. $d=10$. Then the square of (\ref{ukul}) with $k,l = 1,3$
implies an additional relation between monomials, which is denoted
by $[=]$ in the table and subtracts $1$ in evaluating the number
$N_k^{circ}$.

If (\ref{ukul}) is not taken into account, then the numbers $\tilde
N^{circ}_k$ are defined by the same relation (\ref{tildeNvsN}), only
now $N^{circ}_k$ enter the r.h.s. instead of $N_k$. However,
(\ref{tildeNvsN}) requires an additional modification, which takes
(\ref{ukul}) into account, and this modification is {\it linear} in
(\ref{ukul}). For example, from $\tilde N_6^{circ} = 4$ one still
needs to subtract $1$, associated with potential, but vanishing due
to (\ref{ukul}) contribution $-\dot u_\mu u_{13} + \dot{\ddot u}_\mu
u_{11} = \Big(-\dot u_\mu \dot{\ddot u}_\nu + \dot u_\nu \dot{\ddot
u}_\mu\Big) \dot u^\nu = 0$ to $\xi^{(8)}_\mu$. Similarly, from
$\tilde N_8^{circ} = 7$ one needs to subtract $3$, associated with
three such structures $0 = \Big(-\dot u_\mu \dot{\ddot u}_\nu + \dot
u_\nu \dot{\ddot u}_\mu\Big) \dot{\ddot u}^\nu = -\dot u_\mu u_{33}
+ \dot{\ddot u}_\mu u_{13}$,\ $0 = \Big(-\dot u_\mu \dot{\ddot
u}_\nu + \dot u_\nu \dot{\ddot u}_\mu\Big) \dot u^\nu u_{11} = -\dot
u_\mu u_{11}u_{13} + \dot{\ddot u}_\mu u_{11}^2$ and $0 = \Big(-\dot
u_\mu \dot{\ddot{\ddot u}}_\nu + \dot u_\nu \dot{\ddot{\ddot
u}}_\mu\Big) \dot u^\nu = -\dot u_\mu u_{15} + \dot{\ddot{\ddot
u}}_\mu u_{11}$.

The table demonstrates that $\tilde N_k^{circ} = N_k^{circ}$ for all
even $d$, at least, till $d=10$, and this justifies the use of
Kosyakov's trick for evaluation of the radiation friction {\it for
the circular motion} for these dimensions.

The generating function \be {\cal N}_1^{circ}(q) = \frac{q^2}{1-q^2}
= (1-q){\cal N}_1(q) \ee Thus, before subtractions, \be {\cal
N}^{circ}(q) = \exp \left(\sum_{k=1}^\infty \frac{1}{k}{\cal
N}_1^{circ}(q)\right) = \prod_{k=1}^\infty \frac{1}{1-q^{2k}} \ee
and, also before subtractions, \be \tilde {\cal N}^{circ}(q) =
\frac{q}{1-q}{\cal N}^{circ}(q) \ee

\section{Conclusion}

Thus, we conclude that the elegant method, successfully used by
B.Kosyakov to evaluate the radiation friction force in $4$ and $6$
space-time dimensions, can not be directly used in general situation
in higher dimensions (though it still works nicely for the circular
motion with constant angular velocity). Therefore, it seems
unavoidable to make full-scale calculations with explicit
regularization and counter-terms in the action, as it has been done
in $6d$ in \cite{regcal}. For first results in this direction beyond
$6$ dimensions, see \cite{Gala}. Among interesting questions,
appearing on this way, is the counterterms dependence on the choice
of regularization (if all counterterms can be varied independently)
and physical interpretation of emerging corrections to the naive
action of relativistic particle \cite{DBS}.

\section*{Acknowledgements}

Our work is partly supported by Russian Federal Nuclear Energy
Agency, by the joint grant 06-01-92059-CE, by NWO project
047.011.2004.026, by INTAS grant 05-1000008-7865, by
ANR-05-BLAN-0029-01 project and by the Russian President's Grant of
Support for the Scientific Schools NSh-8004.2006.2, by RFBR grants
07-02-00878 (A.Mir.) and 07-02-00645 (A.Mor.).


\begin{thebibliography}{12}


\bibitem{Kos} B.Kosyakov, Theor.Math.Phys. {\bf 119} (1999) 493-505,
hep-th/0207217

\bibitem{grarad} Evaluation of $W_\mu$ for
electromagnetic radiation is a straightforward student-level
exercise with Li\'enard-Wiechert potentials, at least for even
space-time dimension $d$. The case $d=4$ is considered in all
possible textbooks, for example, in

L.Landau and E.Lifshitz, {\sl The Classical Theory of Fields
(Course of Theoretical Physics, Volume 2)}\\
J.D.Jackson, {\sl Classical Electrodynamics,} Wiley, New York,
1975\\

For additional pecularities of generalization to gravitational
radiation, see

 I.Khriplovich, {\sl General Relativity} (in Russian), Izhevsk,
 2001\\

For a textbook, presenting a through discussion of all the details,
also in the case of $d=6$, but only for electromagnetic radiation,
see

 B.Kosyakov, {\sl Introduction to the Classical Theory of
Particles and Fields}, Springer, 2007\\

Radiation in higher even dimensions is considered in the set of
papers:

paper \cite{Kos}\\
D.Galtsov, Phys.Rev. {\bf D66} (2002) 025016, hep-th/0112110 \\
paper \cite{regcal}\\
V.Cardoso, O.Dias and J.Lemos,
Phys.Rev. {\bf D67} (2003) 064026, hep-th/0212168\\
M.Gurses and O.Sarioglu, {\it Class.Quant.Grav.} {\bf 19} (2002)
4249; {\bf 20} (2003) 351; hep-th/0303078\\
Yu.Yaremko, J.Phys. {\bf A37} (2004) 1079-1091\\
B.Koch and M.Bleicher, hep-th/0512353\\
P.Krtous and J.Podolsky,
Class.Quant.Grav. {\bf 23} (2006) 1603-1616, gr-qc/0602007\\
H.A.Morales-T\'ecotl, O.Pedraza and L.O.Pimentel, physics/0611241\\
A.Mironov and A.Morozov, Pisma Zh.Eksp.Teor.Fiz. {\bf 85} (2007)
9-14 (JETP letters, {\bf 85} (2007) 6-11), hep-ph/0612074\\
V.Cardoso, M.Cavaglia and J.-Q.Guo, hep-th/0702138\\

For exhaustive
set of formulas for any even $d$ see \cite{MM2}. This paper contains
expression also for radiated fields of all spins $s$, but or $s>1$
the radiation problem for a point-like source is not well-defined,
so these formulas are not of direct applicability -- contribution of
radiation from other sources, standing behind the force $f_\mu$ in
(\ref{eqm}) should be also included.

\bibitem{MM2} A.Mironov and A.Morozov, arXiv:hep-th/0703097 (to
appear in Theor.Math.Phys.)

\bibitem{regcal} P.Kazinsky, S.Lyakhovich and A.Sharapov, Phys.Rev. {\bf D66} (2002)
025017, hep-th/0201046

\bibitem{Gala} D.Galakhov, arXive:0710.5688 (hep-th)

\bibitem{DBS} P.Dunin-Barkovsky and A.Sleptsov, {\it work in progress}







\end{thebibliography}
\end{document}